\documentclass[12pt]{spieman}

\usepackage{amsfonts}
\usepackage{amsmath}

\usepackage{amssymb}
\usepackage{stackrel}
\usepackage{booktabs}
\usepackage{tabularx}
\usepackage{mathtools}
\usepackage{pgfplots}
\usetikzlibrary{spy}
\usepackage{geometry}
\usepackage{pdflscape}
\usepackage{enumerate}
\usepackage[ruled]{algorithm2e}
\usepackage{esvect}
\usepackage[figuresright]{rotating}
\usepackage{footnote}
\usepackage{listings}
\usepackage{xcolor}
\usepackage{enumitem}
\usepackage{tocloft}
\usepackage{textcomp}
\usepackage{braket}
\usepackage{nicefrac}
\usepackage{soul}
\usepackage{nccmath}
\usepackage{bm}
\usepackage{import}
\usepackage{dpfloat}
\usepackage{nicefrac}
\usepackage{subcaption}
\usepackage{float}
\usepackage{afterpage}
\usepackage{multirow}
\usepackage{siunitx}
\usepackage{graphicx}
\usepackage{tikz}
\usepackage[T1]{fontenc}
\usepackage{xr-hyper}
\usepackage{changes}
\usepackage{hyperref}
\usepackage{pdfcomment}
\DeclarePairedDelimiter\abs{\lvert}{\rvert}%

\newcounter{resultc}

\makesavenoteenv{tabular}

\graphicspath{{Figs/}}

\title{Linear time algorithm for phase sensitive holography}

\author[a,*]{Peter J. Christopher}
\author[a]{Ralf Mouthaan}
\author[a]{Miguel El Guendy}
\author[a]{Timothy D. Wilkinson}
\affil[a]{Centre of Molecular Materials, Photonics and Electronics, University of Cambridge}

\cftpagenumbersoff{figure}
\cftpagenumbersoff{table} 

\begin{document} 
    \maketitle

    \begin{abstract}
    		
        Holographic search algorithms such as direct search and simulated annealing allow high-quality holograms to be generated at the expense of long execution times. This is due to single iteration computational costs of $O(N_x N_y)$ and number of required iterations of order $O(N_x N_y)$, where $N_x$ and $N_y$ are the image dimensions. This gives a combined performance of order $O(N_x^2 N_y^2)$. 
            
        In this paper we use a novel technique to reduce the iteration cost down to $O(1)$ for phase-sensitive computer generated holograms giving a final algorithmic performance of $O(N_x N_y)$. We do this by reformulating the mean-squared error metric to allow it to be calculated from the diffraction field rather than requiring a forward transform step. For a $1024\times 1024$ pixel test images this gave us a $\approx 50,000\times$ speed-up when compared with traditional direct search with little additional complexity. 
    				
        When applied to phase-modulating or amplitude-modulating devices the proposed algorithm converges on a global minimum mean squared error in $O(N_x N_y)$ time. By comparison, most extant algorithms do not guarantee a global minimum is obtained and those that do have a computational complexity of at least $O(N_x^2 N_y^2)$ with the naive algorithm being $O((N_xN_y)!)$.
    \end{abstract}
    
    
    {\noindent \footnotesize\textbf{*}Peter J. Christopher,  \linkable{pjc209@cam.ac.uk} }
    
    \begin{spacing}{2}   
    
    \section{Introduction}
    
    Holographic technology has developed significantly since its invention in 1948 by Dennis Gabor~\cite{gabor1948new}. Conventional holography, developed since then, captures the interference pattern between a coherent light source and the light scattered off an object, onto a photographic plate~\cite{leith1962reconstructed}. A 3-dimensional image of the object is then reconstructed when the photographic plate is exposed to a coherent light source.
    
    The 1980’s saw a breakthrough in holographic technology with the introduction of computer-generated holography (CGH). Improvements in computer processing power and the availability of computer-controlled spatial light modulators (SLMs) gave users more flexible approaches to modulating the spatial profile of an incident beam. In other words, the SLM enabled the flexible configuration of a hologram, something not possible using photographic plates. Advancements in this technology has revolutionized the display industry with it being applied in virtual (VR) and augmented reality (AR) systems~\cite{maimone2017holographic, widjanarko2020clearing, svoboda2013holographic}. In turn positively influencing the wider information and education industries~\cite{yuen2011augmented, lee2012augmented}, as well as healthcare~\cite{pessaux2015towards} and manufacturing~\cite{nee2012augmented}. Holographic technology has also been used in lithography~\cite{campbell2000fabrication} and optical tweezing~\cite{grieve2009hands}.
    
    \begin{figure}
        \centering
        {\includegraphics[trim={0 0.5cm 0 0},width=0.8\linewidth,page=5]{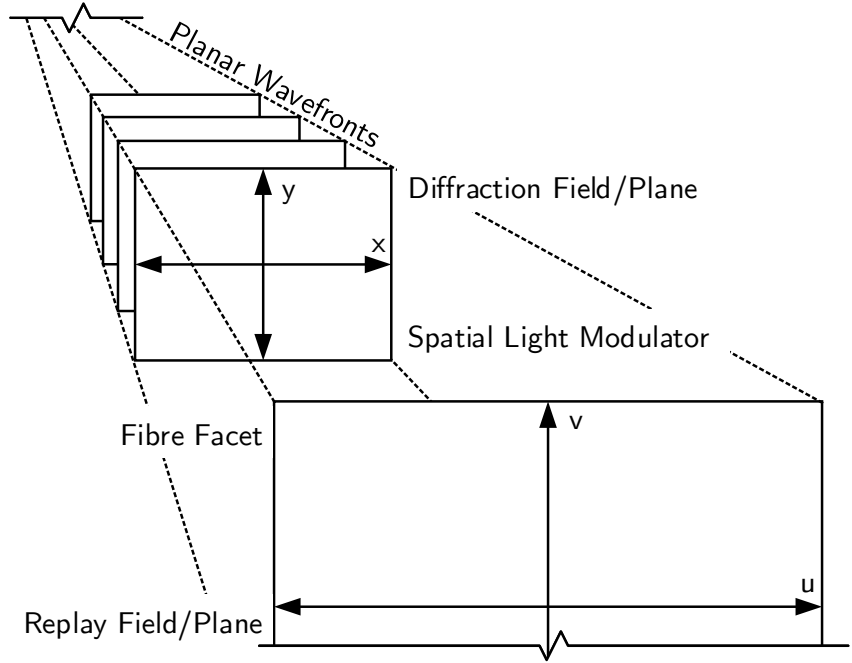}}
        \caption{Coordinate systems used in this work.}
        \label{fig:coords}
    \end{figure}
    
    In modern two-dimensional holography systems, a spatial light modulator (SLM) is used to modulate the profile of a coherent light beam. In the simplest configuration, an SLM is placed at the back focal plane of a lens with the aim of creating a desired light field at the front focal plane of the lens, as shown in Fig. \ref{fig:coords}. The back focal plane is termed the \textit{diffraction field} $H$ and the front focal plane is known as the \textit{replay field} $R$, with the light fields in the two planes related by a Fourier transform $\mathcal{F}$ such that $R = \mathcal{F}\{H\}$. The aforementioned holograms are known as \textit{Fraunhofer holograms}, and it is also possible to project light fields onto planes offset from the front focal plane, which are then known as \textit{Fresnel holograms}.
    
    An SLM is a pixellated device, and as such it is intuitive to represent the diffraction field as discrete pixels. Similarly, the replay field can be represented by discrete pixels and the Fraunhofer transform relationship between the two planes can then be represented by two discrete Fourier transform relationships where the diffraction field coordinates are represented by $x$ and $y$ and the replay field coordinates are represented by $u$ and $v$. $N_x$ and $N_y$ denote the size of the diffraction and replay fields along the horizontal and vertical axes respectively. This expression assumes the SLM is illuminated with a plane wave of uniform intensity, and that the pixels have a fill factor of 100\%.
    
    \begin{align}
    R_{u,v} & = \frac{1}{\sqrt{N_xN_y}}\sum_{x=0}^{N_x-1}\sum_{y=0}^{N_y-1} H_{x,y}e^{-2\pi i \left(\frac{u x}{N_x} + \frac{v y}{N_y}\right)} \label{fouriertrans2d5c} \\
    H_{x,y} & = \frac{1}{\sqrt{N_xN_y}}\sum_{u=0}^{N_x-1}\sum_{v=0}^{N_y-1} R_{u,v}e^{2\pi i \left(\frac{u x}{N_x} + \frac{v y}{N_y}\right)}  \label{fouriertrans2d5d} 
    \end{align}
    
    SLMs allow either the phase or amplitude of the incident light to be modulated, but not both~\cite{STTM}. Additionally, it is often the case that SLMs are digital devices and that only discrete modulation levels can be used. Projecting the desired replay field, known as the \textit{target field} $T$, then corresponds to finding a diffraction field subject to these constraints that minimises some error metric \cite{SSQ}. In this case the phase-sensitive mean-squared error (MSE) shall be used.
    
    \begin{equation} \label{msePS}
    \text{Error}(T, R)= \frac{1}{N_x N_y}\sum_{u=0}^{N_x-1}\sum_{v=0}^{N_y-1} \abs{T_{u,v} -  R_{u,v}}^2
    \end{equation}
    
    The task of finding a computer generated hologram becomes equivalent to minimising this error metric. A variety of techniques have been developed to address this and one family of algorithms is briefly described in Section \ref{sec:HSAs}. These algorithms require repeated Fourier transforms, the evaluation of which is computationally expensive. This paper lays out an alternative approach to generating complex-valued (i.e. phase-sensitive) light fields that only requires a single transform to be used, after which the hologram can be determined using computationally inexpensive update steps. The fundamentals of this approach are laid out in Section \ref{sec:LT_BasicPremise} and is incorporated into a search algorithm in Section \ref{sec:LT_SearchAlgorithm}. Section \ref{sec:LT_IndependentPixels} discusses how this approach lends itself to massive parallelisation. Next, more realistic scenarios are considered, with conclusions being drawn for commercially available SLMs in Section \ref{sec:LT_SLMs} and Fresnel holograms being considered in Section \ref{sec:LT_Fresnel}. The algorithm is modified to account for a region of interest in Section \ref{sec:LT_RoI}, which allows a much higher fidelity replay field to be projected. The approach described requires several orders of magnitude less computing power, but still yields replay fields of the highest quality.
    
    \section{Established Holographic Search Algorithms} \label{sec:HSAs}
    
\begin{figure}
        \centering
        {\includegraphics[trim={0 0 0 0},width=\linewidth,page=1]{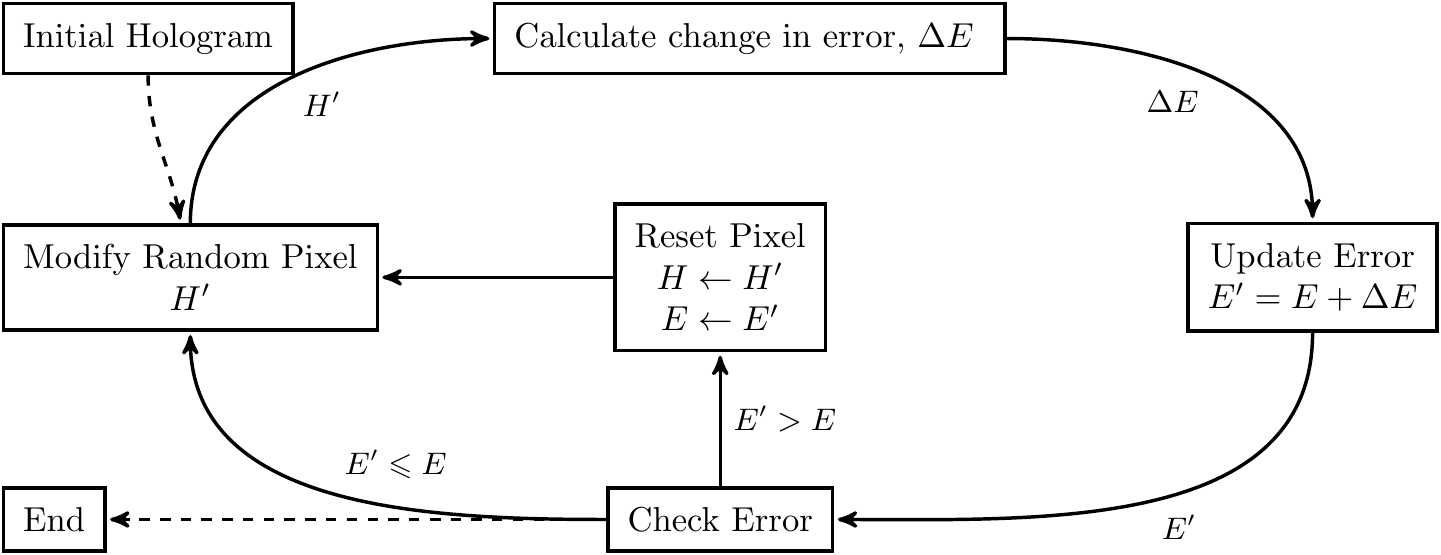}}
        \caption{The direct search algorithm.}
        \label{fig:AlgorithmDS}
    \end{figure}

    A widely used family of algorithms for phase sensitive replay fields are the holographic search algorithms (HSAs), of which the most famous is perhaps direct search (DS) \cite{Clark1996, Jennison1989a, Jennison1989b, Liu2019, Seldowitz1987, KANG2019312}. Broadly speaking, these algorithms proceed by changing a pixel value and evaluating whether the error metric has improved. If the error metric has improved the pixel change is adopted, else the pixel change is rejected. This process is illustrated in Fig. \ref{fig:AlgorithmDS}. 
    
    A second algorithm in this family is simulated annealing (SA) \cite{Carpenter2010, Kirkpatrick1983, Kirk1992, Dames1991, Yang2009}, which sometimes adopts pixel changes that do not improve the error metric in an effort to avoid local minima. HSAs are guaranteed to converge, but can converge extremely slowly and often to local rather than global minimum. Millions of iterations can be required before these algorithms have fully converged. This can be prohibitive if a full FFT is required at each iteration (complexity $O(N_x N_y log(N_x N_y))$). Alternatively, evaluation of the full FFT can be avoided by using an update step that exploits the fact that only a single pixel is updated at a time. This gives an update step with complexity proportional to $O(N_x N_y)$ which is a marked improvement, but can still give long run times for even medium-sized images as the complete algorithm will still run in $O(N_x^2 N_y^2)$. The authors have recently introduced several new holographic search algorithms that exploit geometric arguments to obtain significantly faster convergence \cite{HPS1,HPS2,SortedPixelSelection,AmplitudeLevels}, but these too can still be computationally expensive to run.
    
    \section{Search in linear time}
    
    \subsection{Basic Premise} \label{sec:LT_BasicPremise}
    
    For our initial investigation we shall show that using known properties of the Fourier transform we can significantly reduce the computation required for generating phase sensitive holograms. Note that we are only considering Fraunhofer holograms without a region of interest (RoI), i.e. the entire replay field is to be optimised. We shall extend our analysis to Fresnel holograms and refine our analysis to include an RoI later in this paper. 
    
    The Fourier transform operation obeys Parseval's theorem, reproduced in Eq. \ref{parseval}, where $A = \mathcal{F}(a)$, $B = \mathcal{F}(b)$, and an overline represents the complex conjugate. Parseval's theorem corresponds to energy conservation between the diffraction and replay field planes, which is the reason behind the the $\nicefrac{1}{\sqrt{N_xN_y}}$ term in Eqs.~\ref{fouriertrans2d5c} and \ref{fouriertrans2d5d}. 
    
    \begin{equation} \label{parseval}
    \sum_{x=0}^{N_x-1}\sum_{y=0}^{N_y-1} a_{x,y}\overline{b_{x,y}} = \sum_{u=0}^{N_x-1}\sum_{v=0}^{N_y-1} A_{u,v}\overline{B_{u,v}}
    \end{equation}
    
    The relationship between $H_{x,y}$ and $R_{u,v}$ has previously been defined as a Fourier transform. Similarly, a new field $G_{x,y}$ is defined which corresponds to the inverse Fourier transform of the $T_{u,v}$. In effect, $G_{x,y}$ represents the diffraction field counterpart of the target replay field.
    
    \begin{equation} \label{targets}
    G \stackrel[\mathcal{F}^{-1}]{\mathcal{F}}{\rightleftarrows} T, \quad H \stackrel[\mathcal{F}^{-1}]{\mathcal{F}}{\rightleftarrows} R
    \end{equation}
    
    These definitions can be used with Parseval's theorem to obtain a new expression for the MSE metric.
    
    \begin{align} \label{msePS2}
    E_{MSE} & = \frac{1}{N_x N_y}\sum_{u=0}^{N_x-1}\sum_{v=0}^{N_y-1} \abs{T_{u,v} -  R_{u,v}}^2 \nonumber                                     \\
    & = \frac{1}{N_x N_y}\sum_{x=0}^{N_x-1}\sum_{y=0}^{N_y-1} (T_{u,v} -  R_{u,v})(\overline{T_{u,v}} -  \overline{R_{u,v}}) \nonumber \\
    & = \frac{1}{N_x N_y}\sum_{x=0}^{N_x-1}\sum_{y=0}^{N_y-1} (G_{x,y} -  H_{x,y})(\overline{G_{x,y}} -  \overline{H_{x,y}}) \nonumber \\
    & = \frac{1}{N_x N_y}\sum_{x=0}^{N_x-1}\sum_{y=0}^{N_y-1} \abs{G_{x,y} -  H_{x,y}}^2                                               
    \end{align}
    
    The key innovation of this paper is to observe that this allows us to determine the value of $E_{MSE}$ on the diffraction field side of the transform from $G_{x,y}$ and $H_{x,y}$, and that this avoids the need for repeatedly projecting changes to the replay fields side to calculate the MSE. If we know the original MSE, then the effect of any change can be determined in $O(1)$ rather than the $O(N_x N_y)$ time required for a calculation on the replay field side.
    
    \refstepcounter{resultc}\textbf{Result \theresultc}
    Mean squared error calculation for any phase sensitive Fraunhofer hologram can be done in the diffraction plane.
    
    \subsection{A linear-time holographic search algorithm}\label{sec:LT_SearchAlgorithm}
    
\begin{figure}
        \centering
        {\includegraphics[trim={0 0 0 0},width=1.0\linewidth,page=1]{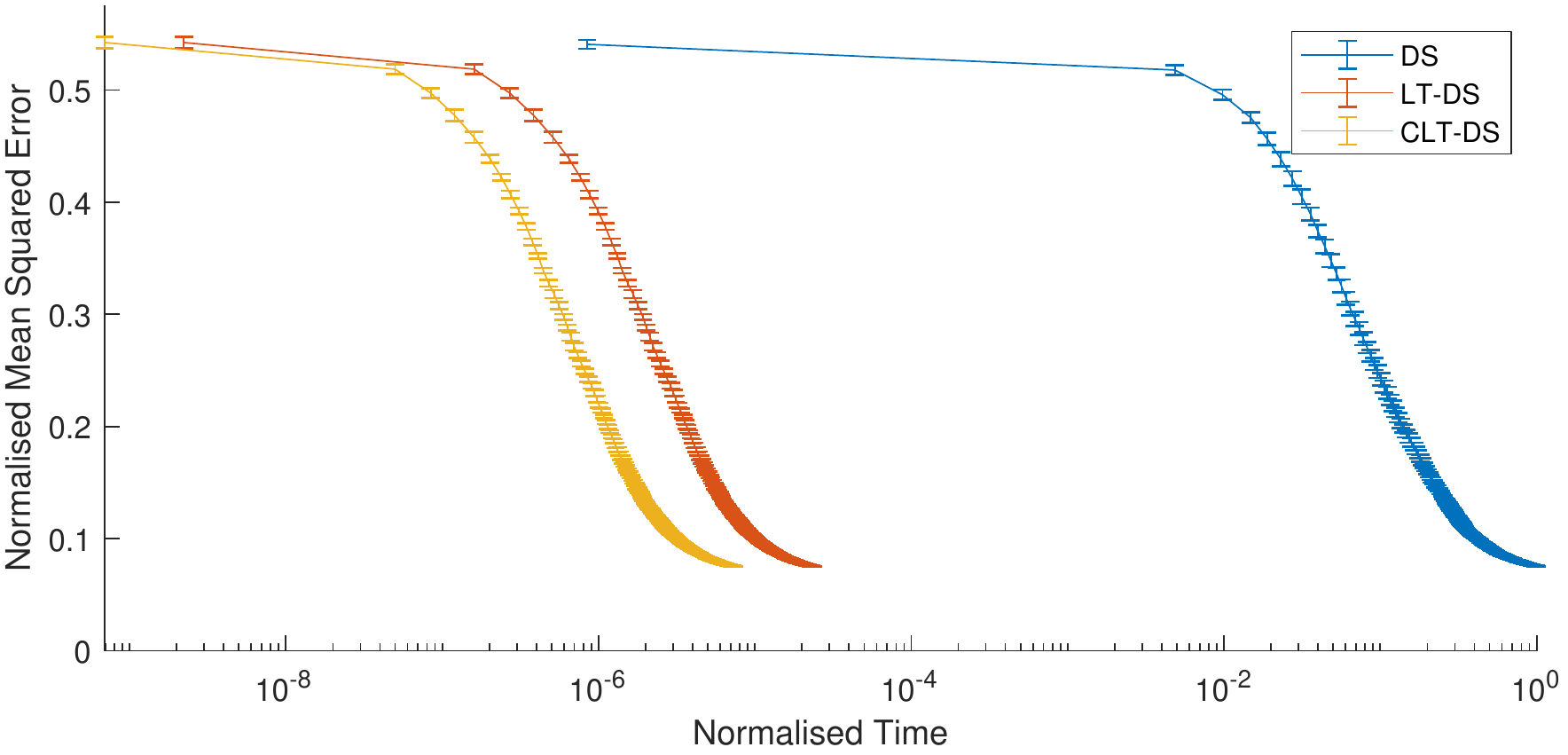}}
        \caption{Performance of direct search and linear time direct search for a simulated $1024\times 1024$ pixel $2^8$ phase level spatial light modulator. Target amplitudes are given by the \textit{Mandrill} test image and target phases are given by the \textit{Peppers} test image as shown in Figure \ref{fig:TestImages}.}
        \label{fig:LinearTime1}
    \end{figure}

\begin{figure}[htbp]
        \centering
        {\includegraphics[trim={0 0 0 0},width=0.66\linewidth,page=1]{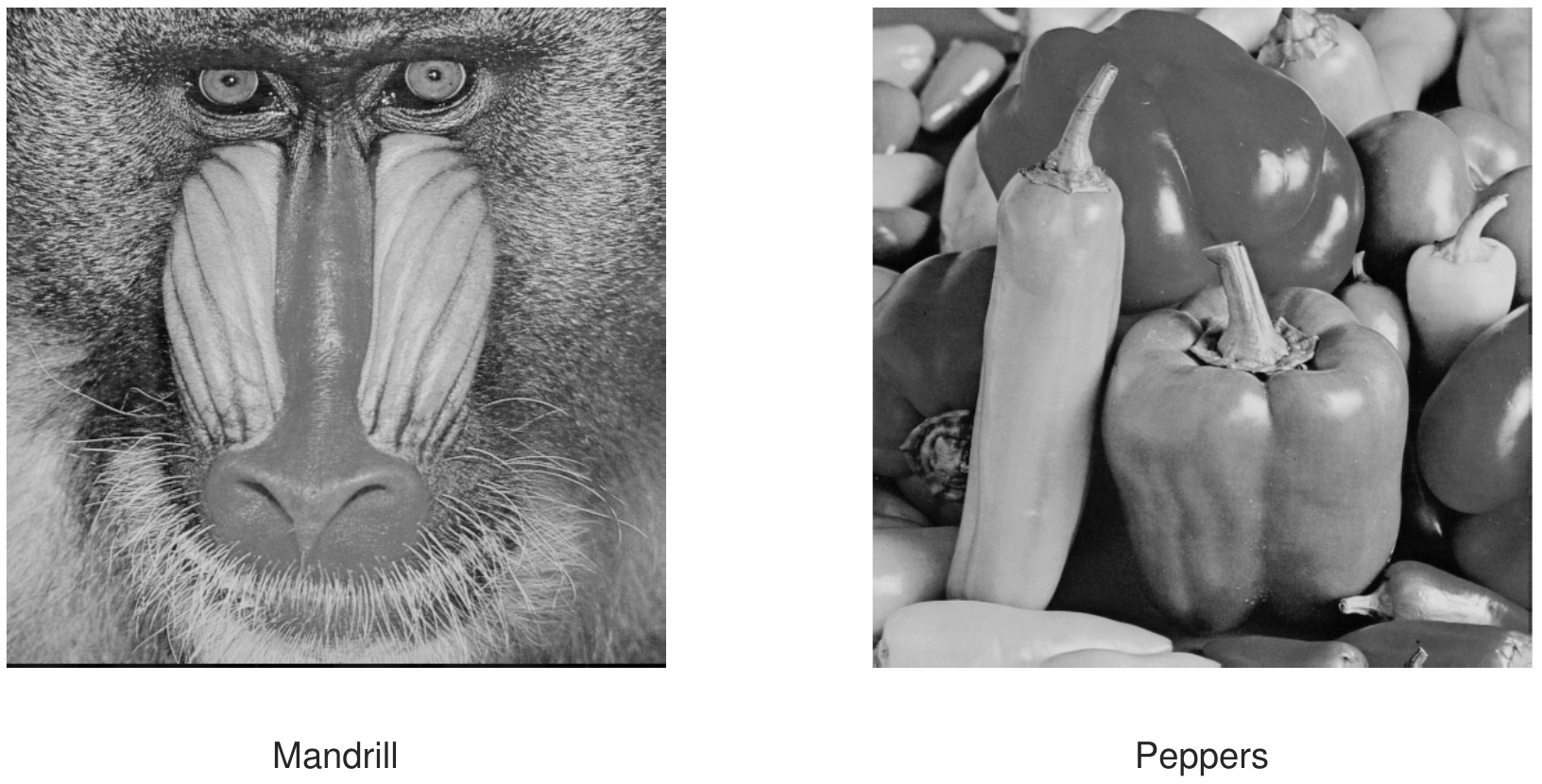}}
        \caption{The two test images used.}
        \label{fig:TestImages}
    \end{figure}

    Crucially, the calculation of $G_{x,y}$ needs to only be done once - before the hologram calculation commences - in other words, there is no longer a need for repeated Fourier transform evaluations at each iteration. While it may appear obvious that Eq.~\ref{parseval} necessitates that Eq.~\ref{msePS} and Eq.~\ref{msePS2} are equivalent, we are unaware of this result having been used previously for hologram generation. Importantly, if we know $E_{MSE}(G,H)$, changing a single pixel in $H$ at coordinates $x$, $y$ allows us to write an expression for the new error 
    
    \begin{equation} \label{msePS3}
    \Delta E_{MSE}(G,H) = \abs{G_{x,y} -  H_{x,y}-\Delta H_{x,y}}^2-\abs{G_{x,y} -  H_{x,y}}^2
    \end{equation}
    
    which runs in constant $O(1)$ time whereas on the replay side the update runs in $O(N_xN_y)$ time. This error calculation can be incorporated into the direct search algorithm (Fig. \ref{fig:AlgorithmDS}) to give linear time direct search (LT-DS). Running the LT-DS algorithm gives the performance graph shown in Figure \ref{fig:LinearTime1}. Target amplitudes are given by the \textit{Mandrill} test image and target phases are given by the \textit{Peppers} test image as shown in Figure \ref{fig:TestImages}. With $1024\times 1024$ pixel test images this gave a $\approx 50,000\times$ speed up for the DS algorithm. Similar results are seen for simulated annealing. Due to the amplitude and phase constraint on the target, however, convergent reconstruction quality is extremely poor. This is traditionally solved by using a region of interest, a topic we return to in Section \ref{sec:LT_RoI}.
    
    It is important to note that, provided the random number generators have the same seed, the hologram given by LT-DS is identical in every way to the hologram provided by DS. The only difference is the Fourier plane on which calculation occurs and the resulting orders of magnitude speed-up. Also worthy of note is that we have normalised the values of the hologram here to give a mean of unit energy per pixel on SLM and replay field sides, with a resulting normalisation effect on the MSE.
    
    \refstepcounter{resultc}\textbf{Result \theresultc}
    The change in mean squared error of a phase sensitive hologram due to a single pixel change can be found in constant  $O(1)$ time.
    
    \subsection{Effect of independence}\label{sec:LT_IndependentPixels}
    
    Section \ref{sec:LT_SearchAlgorithm} used Eq.~\ref{msePS3} to reduce the computation required for DS but maintained the use of the search approach. There are cases, such as when an RoI is taken into account (Section \ref{sec:LT_RoI}), where a search approach is necessary, but for the RoI-free case discussed here we do not actually need to use search syntax at all. Instead, we notice that the effect on the MSE of changing a single pixel is independent of the other pixels. This means that we can actually remove the search element altogether, instead independently assigning values to each individual pixel. This is important as it allows us to parallelise the algorithm for multi-core devices. The performance improvement obtained in this way over the sequential version is also shown in Fig. \ref{fig:LinearTime1} and we have termed it concurrent LT-DS or CLT-DS. The workstation used had a Intel\textsuperscript{\textregistered} i7-9900K CPU, overclocked to 5.0GHz with 64GB of 4000MHz DDR4 ram and a RTX 2080TI GPU.
    
    \refstepcounter{resultc}\textbf{Result \theresultc}
    The change in mean squared error of a far-field phase sensitive hologram due to a single pixel change is independent of the effect of other pixels.
    
    \subsection{Realistic SLM constraints}\label{sec:LT_SLMs}
    
    The form of Eq. \ref{msePS2} is a linear minimisation problem and is solvable analytically for a range of modulation regimes. This dependency on the properties of the modulator requires us to investigate the case of phase and amplitude modulating devices separately.
    
    \subsubsection{Phase modulating} \label{phasemod}
    
    If we assume a phase modulating device where $H_{x,y}$ is confined to the complex circle with magnitude given by the incident illumination $I_{x,y}$ then we can reformat Eq.~\ref{msePS3} 
    
    \begin{align} \label{hpsphase1}
    \text{minimise} \quad & \sum_{x=0}^{N_x-1}\sum_{y=0}^{N_y-1}\abs{G_{x,y} -  H_{x,y}}^2 \nonumber \\
    \rightarrow \quad     & \Phi_H = \Phi_G                                                          
    \end{align}
    
    where $\Phi_G$ and $\Phi_H$ correspond to the phase vectors of $G$ and $H$.
    
    \refstepcounter{resultc}\textbf{Result \theresultc}
    When aberration and replay field RoIs are neglected, the lowest possible mean square error is achieved for a far-field phase hologram when the phase is equal to the inverse transform of the target replay. 
    
    \subsubsection{Amplitude modulating} \label{ampmod}
    
    If we assume an amplitude modulating device where $H_{x,y}$ is assumed to be confined to $\abs{H_{x,y}}\geqslant0$ and $\Phi_H=0$ then  we can reformat Eq.~\ref{msePS3} 
    
    \begin{equation} \label{hpsamplitude1}
    \text{minimise} \quad \sum_{x=0}^{N_x-1}\sum_{y=0}^{N_y-1}\abs{G_{x,y} -  H_{x,y}}^2 \quad \rightarrow \quad H = \Re(G)
    \end{equation}
    
    \textbf{\refstepcounter{resultc}\textbf{Result \theresultc}}
    When aberration and replay field RoIs are neglected, the lowest possible mean square error is achieved for a far-field amplitude hologram when the SLM amplitude is equal to the real part of the inverse transform of the target replay.
    
    \subsection{Fresnel holograms, aberration correction and 3D} \label{sec:LT_Fresnel}
    
    The Fresnel transform used for generating Fresnel holograms is equivalent to the Fourier transform with the addition of a \textit{quadratic phase factor} as in
    
    \begin{equation} 
    R_{u,v} = \underset{\scriptscriptstyle \text{Fresnel}}{\mathcal{F}}\{H_{x,y}\} = \underset{\scriptscriptstyle \text{Fourier}}{\mathcal{F}}\{H_{x,y}\Phi_\text{Fresnel}\}
    \end{equation}
    
    where $\Phi_\text{Fresnel} = \exp{\frac{i\pi}{\lambda z}(x^2 + y^2)}$. It can be seen that the Parseval theorem remains applicable here; Eqs.~\ref{msePS} and \ref{msePS2} remain equivalent and the results of Sections~\ref{phasemod} and \ref{ampmod} remain valid with the addition of an additional phase term.
    
    In fact, for any input phase term dependent only on $x$ and $y$ we can assert the equivalence of Eq.~\ref{msePS} and Eq.~\ref{msePS2}. This includes the family of Seidel aberrations.
    
    While we discuss the linear-time algorithm here in the context of 2D holograms, it is equally applicable to 3D holograms generated by means of \textit{Fresnel slices} or the layer based technique. 
    
    \section{Incorporating a region of interest}\label{sec:LT_RoI}
    
    The reconstruction quality obtained for complex-valued target fields using the techniques above is often extremely poor, but this is not due to the choice of algorithm. Instead, this is because the problem is over-constrained. One solution that is widely used is to only require a portion of the replay field to match the target image, with the remainder of the replay field being free to take on any value. Mathematically we can define a region of interest mask $M_{u,v}$ where $M_{u,v}=1$ in the region of interest and $M_{u,v}=0$ otherwise. We then we can write mean squared error as
    
    \begin{equation} \label{msePS5}
    E_{MSE}(T, R)=\frac{1}{N_x N_y}\sum_{u=0}^{N_x-1}\sum_{v=0}^{N_y-1} M_{u,v}\abs{T_{u,v} -  R_{u,v}}^2
    \end{equation}
    
    Unfortunately we can no longer use Eq.~\ref{parseval} in order to move this to the SLM side, as Parseval's theorem only holds true if all of space is considered, instead of only a subregion of space. 
    
    We present here an alternative technique for incorporating an RoI into a linear time algorithm. We can rewrite Eq.~\ref{msePS5} to give the following 
    
    \begin{alignat}{2} \label{RF1}
    E_{MSE} & = \frac{1}{N_x N_y}\sum_{u=0}^{N_x-1}\sum_{v=0}^{N_y-1} \enspace&& \abs{M_{u,v}T_{u,v} -  M_{u,v}R_{u,v}}^2 \nonumber\\
    & = \frac{1}{N_x N_y}\sum_{u=0}^{N_x-1}\sum_{v=0}^{N_y-1} \enspace&& M_{u,v}T_{u,v}\overline{M_{u,v}T_{u,v}} - M_{u,v}T_{u,v}\overline{M_{u,v}R_{u,v}} -	\nonumber\\
    & &&\overline{M_{u,v}T_{u,v}}M_{u,v}R_{u,v} + M_{u,v}R_{u,v}\overline{M_{u,v}R_{u,v}} \nonumber\\
    & = \frac{1}{N_x N_y}\sum_{x=0}^{N_x-1}\sum_{y=0}^{N_y-1} \enspace&& F_{x,y}\overline{F_{x,y}} - F_{x,y}\overline{(L*H)_{x,y}} - \nonumber\\
    & &&\overline{F_{x,y}}(L*H)_{x,y}+(L*H)_{x,y}\overline{(L*H)_{x,y}}\nonumber\\
    & = \frac{1}{N_x N_y}\sum_{x=0}^{N_x-1}\sum_{y=0}^{N_y-1} \enspace&& F_{x,y}\overline{F_{x,y}} - F_{x,y}\overline{K_{x,y}} - \overline{F_{x,y}}K_{x,y}+K_{x,y}\overline{K_{x,y}} 
    \end{alignat}
    
    where  `$*$' denotes convolution, `$\cdot$' denotes the Hadamard or `dot' product and 
    
    \begin{equation} \label{targets2}
    L \stackrel[\mathcal{F}^{-1}]{\mathcal{F}}{\rightleftarrows} M , \quad
    F \stackrel[\mathcal{F}^{-1}]{\mathcal{F}}{\rightleftarrows} M\cdot T  , \quad
    K \stackrel[\mathcal{F}^{-1}]{\mathcal{F}}{\rightleftarrows} M\cdot R \nonumber
    \end{equation}
    
    $F_{x,y}$ behaves similarly to our previous study and single pixel updates can be determined in $O(1)$. $K_{x,y}$ corresponds to a convolution though, and cannot be evaluated as easily. Fortunately, while convolution is an $O(N_x^2N_y^2)$ problem, changing a single pixel of a convolution can be somewhat optimised. The convolution term of Eq.~\ref{RF1} is given for any pixel $x'$, $y'$ by
    
    \begin{equation}
    K_{x',y'} = \sum_{a=0}^{N_x-1}\sum_{b=0}^{N_y-1} L_{a,b}H_{x'-a,y'-b} 
    \end{equation}
    
    Recognising that $L$ is only non-zero for a handful of pixels, this can be calculated in $O(n)$ where $n$ is the number of pixels where $L \neq 0$. 
    
    \begin{equation}
    K_{x',y'} = \sum_{L \neq 0} L_{a,b}H_{x'-a,y'-b} 
    \end{equation}
    
    A change in a single pixel $x$, $y$ of value $\Delta H_{x,y}$ then causes a difference to the convolution at pixel $x'$, $y'$ of
    
    \begin{equation}
    \Delta K_{x',y'} = L_{x'-x,y'-y}\Delta H_{x,y} 
    \end{equation}
    
    Incorporating this back into the MSE equation, the following update step can then be defined.
    
    \begin{equation} \label{RF3}
    \Delta E_{MSE} = \frac{1}{N_x N_y}\sum_{x'=0}^{N_x-1} F_{x,y}\overline{\Delta K_{x,y}} - \overline{F_{x,y}} \Delta K_{x,y}+\Delta K_{x,y}\overline{K_{x,y}}+
    K_{x,y}\overline{\Delta K_{x,y}}+
    \Delta K_{x,y}\overline{\Delta K_{x,y}}
    \end{equation}
    
    This can be incorporated into the DS algorithm of Fig. \ref{fig:AlgorithmDS}. Any given mask $M_{u,v}$ can be given to an arbitrary degree of accuracy by $\mathcal{F}\{L\}$ though in practice if $L$ is non-zero for more than a few points we recommend a change of mask or an alternative approach.
    
\begin{figure}
        \centering
        \begin{tikzpicture}[spy using outlines={circle, magnification=10,connect spies}]
            \node[anchor=south west,inner sep=0] (image) at (0,0) {\includegraphics[trim={0 0 0 0},width=1.0\textwidth,page=1]{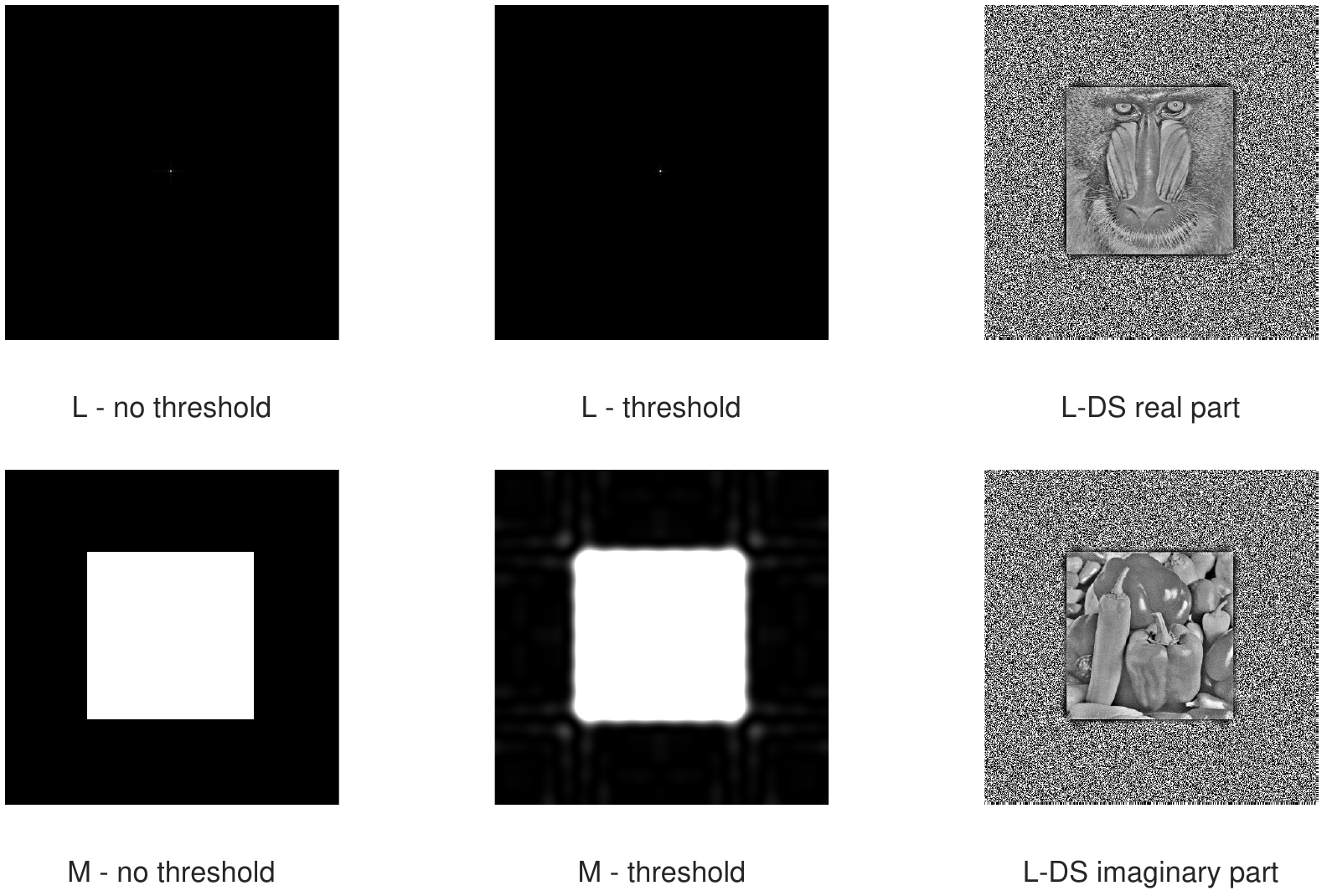}};
            \begin{scope}[x={(image.south east)},y={(image.north west)}]
                \coordinate (coord1) at (.1288,.808);
                \coordinate (coord2) at (.25,.9);
                \coordinate (coord3) at (.4975,.808);
                \coordinate (coord4) at (.62,.9);
                \spy[gray,width=0.15\textwidth,height=0.15\textwidth] on (coord1) in node at (coord2);
                \spy[gray,width=0.15\textwidth,height=0.15\textwidth] on (coord3) in node at (coord4);
            \end{scope}
        \end{tikzpicture}

        \caption{Mask and inverse transform of mask without thresholding (left) and with thresholding (centre) Reconstruction real and imaginary parts for LT-DS are shown right. Target amplitudes are given by the \textit{Mandrill} test image and target phases are given by the \textit{Peppers} test image as shown in Figure \ref{fig:TestImages}.}
        \label{fig:LinearTime6}
    \end{figure}
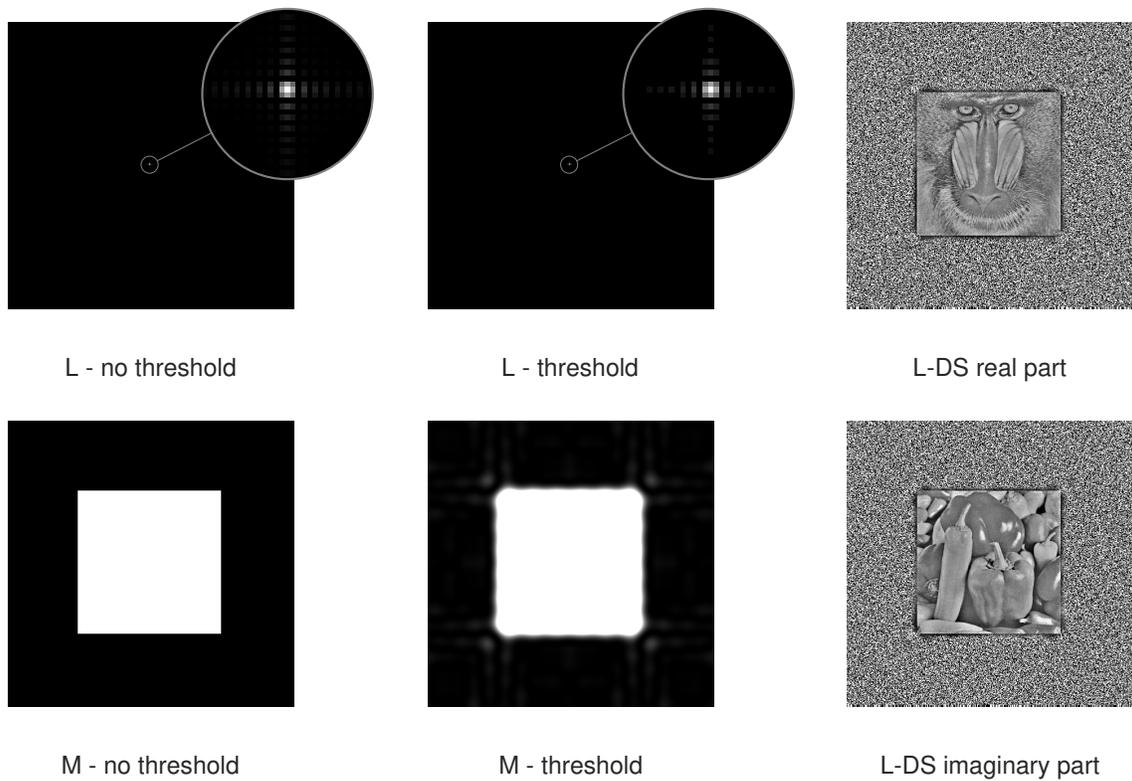

    To demonstrate this in action we take the case of $L$ being non-zero only at a selected 45 points out of a $512\times512$ image. This leads to a mask function similar to Figure~\ref{fig:LinearTime6} with associated figures. 
    
    The quality of the mask in Figure~\ref{fig:LinearTime6} depends on the thresholding value chosen. For many simple masks, over 90\% of the power in the mask can be captured by only a few points in $L$. This corresponds to a slight re-weighting of MSE priorities due to differences in value of $M$.
    
\begin{figure}
        \centering
        {\includegraphics[trim={0 0 0 0},width=1.0\linewidth,page=1]{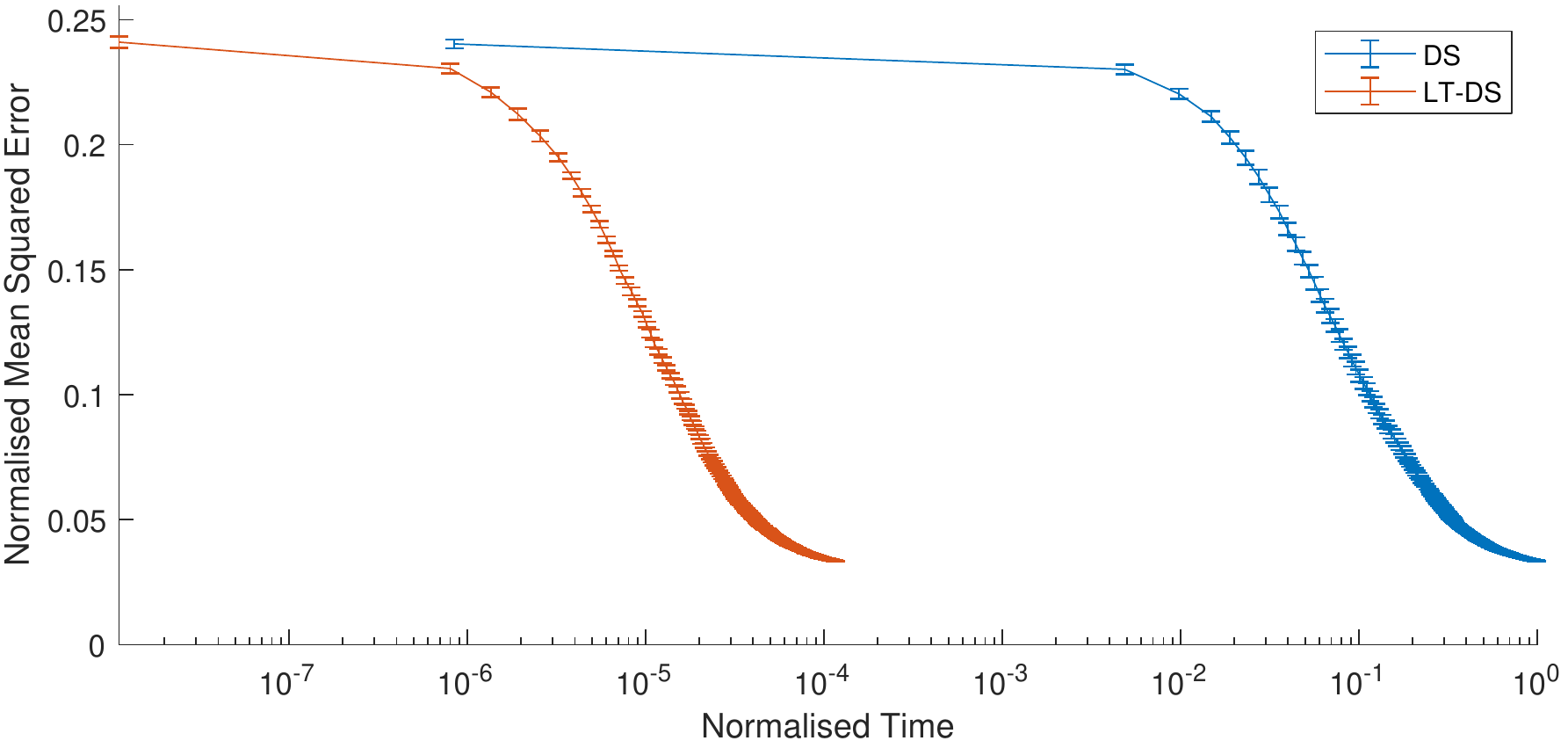}}
        \caption{Performance of direct search and linear time direct searchfor a simulated $1024\times 1024$ pixel $2^8$ phase level spatial light modulator with mask region thresholded at 45 points. Target amplitudes are given by the \textit{Mandrill} test image and target phases are given by the \textit{Peppers} test image as shown in Figure \ref{fig:TestImages}.}
        \label{fig:LinearTime1a}
    \end{figure}

    The performance scales linearly with the number of points in $L$. For the images in Figure~\ref{fig:LinearTime6} with $L$ thresholded to 45 points, we see the performance shown in Figure~\ref{fig:LinearTime1a} with identical normalisation to that in Figure~\ref{fig:LinearTime1}. The speed improvement when compared to Figure~\ref{fig:LinearTime1} is lower, however, due to higher number of calculations per iteration but is still $10,000\times$ faster than the traditional DS approach.
    
    As in Section~\ref{sec:LT_SearchAlgorithm}, the hologram generated using this approach is identical to generating a hologram using DS with mask function $M$ provided the same random number generator seeds are used in both cases. 
    
    \section{Further Work}
    
    The work described so far is applicable in the case where both the phase and amplitude of the replay field are to be controlled. The progress made prompts the obvious question of whether this linear time technique can be applied to phase insensitive holograms where the error is given by
    
    \begin{equation} \label{msePI}
    E_{MSE,pi} = \frac{1}{N_x N_y}\sum_{u=0}^{N_x-1}\sum_{v=0}^{N_y-1} \left[\abs{T_{u,v}} -  \abs{R_{u,v}}\right]^2
    \end{equation}
    
    Clearly this problem is non-linear so a best possible solution is improbable. The authors believe, however, that the techniques of this paper should allow a similar movement of an error metric to the SLM side, but have so far been unable to implement this.
    
    \section{Conclusions}
    
    This paper has presented a new approach to generating holograms for two-dimensional phase sensitive replay fields. The discussed algorithm relies a judicious use of the Parseval's theorem, allowing the phase-sensitive MSE error metric to be calculated from the field in the SLM plane. This allows search algorithms such as SA and DS to run without the need for repeated Fourier transforms, providing a significant acceleration in execution time. Whereas one iteration of a more traditional DS algorithm has a computational cost of $O(N_x N_y)$, iterations of the new proposed implementation have a computation cost as low as $O(1)$. This performance boost is particularly marked for high-definition holograms. For example, with the Tokyo 2020 Olympics being shown in 8k ($7680 \times 4320$) resolution, the expected performance improvement is over 1 million times faster. The algorithm has been presented for Fraunhofer holograms, but has been shown to be equally valid for Fresnel holograms. Conclusions have been drawn for common modulation schemes. An equivalent approach for a phase-insensitive MSE error metric has not been found, but it is felt that further work can address this.
    
    \section*{Funding}
    
    The authors would like to thank the Engineering and Physical Sciences Research Council (EP/L016567/1, EP/T008369/1, EP/L015455/1 and EP/L015455/1) for financial support during the period of this research.
    
    \section*{Disclosures}
    
    The authors declare no conflicts of interest.
    
    \bibliography{references}
    \bibliographystyle{spiejour}
    
\end{spacing}
\end{document}


\maketitle

    \begin{abstract}
        This document contains supplementary material for the journal paper titled "Linear time algorithm for phase sensitive holography".
    \end{abstract}

\keywords{Computer Generated Holography, Holographic Predictive Search, Direct Search, Simulated Annealing, Linear Time}

{\noindent \footnotesize\textbf{*}Peter J. Christopher,  \linkable{pjc209@cam.ac.uk} }

\begin{spacing}{2}   
    
    \section{Introduction}
     
    This work serves as a supplementary companion document for the journal paper titled "Linear time algorithm for phase sensitive holography". In the referenced paper, we introduced a novel approach to generating holograms that led to speed improvements of more than 50,000 times. While the concepts there were simple, the implementation details are slightly more involved.
     
    \section{Matlab Implementation}
    
    \subsection{Fast Fourier Transform}\label{sec:fft}
    
    In order to generate a hologram using Matlab's inbuilt fast Fourier transform (FFT) capabilities, we require an additional shift operation to ensure that the zero-order of our image is centred. This leads to the following familiar functions
    
    \lstinputlisting[style=Matlab-editor]{scripts/ft2.m}
    
    \lstinputlisting[style=Matlab-editor]{scripts/ift2.m}
    
    The \lstinline[style=Matlab-editor]!fftshift! function here serves to swap diagonally opposite quadrants of the image. Notice also the scale factor that ensures that we satisfy Parseval's condition.
    
    \subsection{Convolution}
    
    Matlab does not ship with a 2D circular convolution. Taking account of the shift operation of Section~\ref{sec:fft} we can write
    
    \lstinputlisting[style=Matlab-editor]{scripts/cconv2-slow.m}
    
    Obviously, we can use the convolution theorem to also write
     
    \lstinputlisting[style=Matlab-editor]{scripts/cconv2-fast.m}
    
    \subsection{Indices} \label{indices}
    
    There are two primary roadblocks to implementing the algorithms discussed in the main paper into Matlab:
    
    \begin{enumerate}
        \item Matlab indexes arrays from $1$ rather than from $0$. This leads to many \textit{off-by-one} style errors unless carefully managed.
        \item The \lstinline[style=Matlab-editor]!fftshift! operation means that we often have to rotate our coordinates by $\nicefrac{N_x}{2}$ or $\nicefrac{N_y}{2}$.
    \end{enumerate}

    \subsection{Normalisation}

    In order to ensure conservation of energy, we load and normalise our target image using the following excerpt.
    
    \lstinputlisting[style=Matlab-editor]{scripts/LoadTarget.m}
    
    \subsection{Regions of Interest: Mean Squared Error}
    
    For handling an RoI we used $L$ and $M$
    
    \begin{equation}
    L \stackrel[\mathcal{F}^{-1}]{\mathcal{F}}{\rightleftarrows} M
    \end{equation}
    
    For a given $M$, we can use the following code to select the most significant values of $L$.
     
    \lstinputlisting[style=Matlab-editor]{scripts/BiggestL.m}
    
    We can then use this to generate a hologram as
    
    \lstinputlisting[style=Matlab-editor]{scripts/RoIType2.m}
    
    where the \lstinline[style=Matlab-editor]!mod! statements are to satisfy the constraints of Section 4. "Incorporating a region of interest."
    
    \section{Conclusion}
     
     In this supplementary document we presented Matlab snippets that will allow the reader to reproduce our results from our paper titled: "Linear time algorithm for phase sensitive holography".
     
     \section*{Funding}
     
     The authors would like to thank the Engineering and Physical Sciences Research Council (EP/L016567/1, EP/T008369/1, EP/L015455/1 and EP/L015455/1) for financial support during the period of this research.
     
     \section*{Disclosures}
     
     The authors declare no conflicts of interest.
        
    \bibliography{references}
    \bibliographystyle{spiejour}
    
\end{spacing}